# Transformations in the Time of The Transformer

Peyman Faratin[1,2]    Ray Garcia[3]    Jacomo Corbo[4]
[1] MIT, [2] Robust Links, [3] Buoyant Capital, [4] PhysicsX
peyman@mit.edu, ray@bcap.biz, jacomo@physicsx.ai

## Abstract

Foundation models offer a new opportunity to redesign existing systems and workflows with a new AI first perspective. However, operationalizing this opportunity faces several challenges and tradeoffs. The goal of this article is to offer an organizational framework for making rational choices as enterprises start their transformation journey towards an AI first organization. The choices provided are holistic, intentional and informed while avoiding distractions. The field may appear to be moving fast, but there are core fundamental factors that are relatively more slow moving. We focus on these invariant factors to build the logic of the argument.

## Introduction

There are few moments in history where the chance to reimagine how we function as society presented itself. From agriculture to mobile revolution we have harnessed technologies to redefine ourselves and our environment. Artificial Intelligence (AI) is the latest such force that is enabling us to reimagine again, this time through a new lens of intelligence and adaptation, with profound effects. Flashes of those futures are already manifesting themselves. For instance, in Microsoft's Fiscal year 2024 first quarter earnings call CEO Satya Nadella, when discussing the architecture of Azure cloud services, provided the following insight into what we will refer to as a new AI first enterprise [3]:

*"It is true that the approach we have taken is a full stack approach all the way from whether it's ChatGPT or Bing Chat or all our Copilots, all share the same model. So in some sense, one of the things that we do have is very, very high leverage of the one model that we used, which we trained, and then the one model that we are doing inferencing at scale."* He then goes on to say:

*"The lesson learned from the cloud side is — we're not running a conglomerate of different businesses, it's all one tech stack up and down Microsoft's portfolio, And that, I think, is going to be very important, because that discipline, given what the spend will look like for this AI transition, any business that's not disciplined about their capital spent accruing across all their businesses could run into trouble."*

The goal of this article is to offer an organizational framework to make rational choices as enterprises start their transformation to an AI first enterprise journey, choices that are holistic, intentional and informed while avoiding distractions. We begin by elaborating on an anticipated future state of AI technology, and then justify this future state by describing the necessary evolutionary path AI will take to this end state. Along the way we will review the core technology (Foundation Models, the "one large model" referenced by Nadella in the narrative above), and set the context and the strategic levers AI affords. Finally, we use this framework to provide strategic and operational AI transformation guidelines in locksteps with the anticipated development.

## The Future: Intelligence as a Service (IQaaS)

We believe intelligence will no longer be a scarce good, and will increase in quality and quantity. Intelligence will consist of two goods: that of human level intelligence and a machine generated "super-intelligence", where the latter is defined as new knowledge provisioned autonomously by AI. The AI knowledge will not come from data but through mechanisms such as "self-play", where AI learns new knowledge from many simulations over synthetically generated data.[1] Self-play requires very well defined domains (like board games or video games), thus we expect the supply of the super-intelligence will be constrained to a finite number of domains.

In the future we expect highly capable vertically integrated and closed AI platform providers to compete with one another and with less capable but more open AI platform providers. The competition will be based on costs, data, compute, speed and quality in organized segmented markets. This good will be generated either monolithically, where a single AI provides all information, or by collaboration provided by many federated intelligent systems. Furthermore, this future will enable instantaneous and effortless access to all forms of human and AI intelligence via natural multimodal interfaces, much like electricity that can be accessed across time and space. Enterprises and individuals can simply plugin to this good without building costly special purpose access or application artifacts. Like energy, the supply network will organize itself in a manner that provides frictionless access to intelligence, anywhere, anyhow and any time, all via universal interfaces. As Altman (CEO of OpenAI) comments:

"*Right now, people [say] 'you have this research lab, you have this API, you have the partnership with Microsoft, you have this ChatGPT thing, now there is a GPT store'. But those aren't really our products,"  Altman said. "Those are channels into our one single product, which is intelligence, magic intelligence in the sky. I think that's what we're about.*" Sam Altman ("OpenAI chief seeks new Microsoft funds to build 'superintelligence'", Financial Times, 2023).

We next describe the context and features of this emerging technology and then use this context to describe the potential evolutionary path of the technology to this future end state.

## Foundation Models

Foundational Models (FMs), such as Large Language Models (LLMs) and Diffusion Models (DMs) so called to "*underscore their critically central yet incomplete character*" [1], are ushering in a cambrian moment in AI, a technology that is positioned to become a key entry in modern computing stack, indeed some would argue emergence of an new AI Operating System (AIOS). The core technology of FMs is a neural network called the Transformer [9] that efficiently parallelizes the fundamental vector-matrix operations of a neural network on dedicated hardware such as GPUs.[2] To understand the impact of this technology, we describe the key features of the ecosystem and technology that form the macro context of the FM ecosystem.

---

[1] We witnessed a glimpse of this in 2016 where the world Go champion Sedol competed against AlphaGo, the AI from Google's Deepmind. AlphaGo played a very unusual move on move 37 that clearly rattled Sedol who left the room for 15 minutes.

[2] At time of writing of this article state-space based models such as Mamba [18] are contending to compete with incumbent Transformer models on more efficient compute footprint.

# Context

There are two contextual backgrounds to address before discussing the future state. Firstly, we call *who* does *what* functionality, *where* in the value chain the system architecture; this architecture induces the stakeholders and their industrial organization in the marketplace. The system architecture of AI is currently in transition, being negotiated in the market. We will elaborate on this evolution next. The stakeholders that are emerging in the FM value chain includes:

- Consumers (Enterprise and Retail)
- Developers
- FM platform Providers (OpenAI, Google, Anthropic, Cohere, Meta, Mistral, …)
- Data Providers (open Internet)
- Infrastructure as a Service (IaaS) Providers (Azure/GCP/AWS)
- Hardware Providers (NVIDIA, Intel, ARM, AWS,...)

At the time of writing of this article there are approximately nine FM platform providers, entities that train and maintain FMs (OpenAI/Microsoft, Google, Meta, Cohere, Inflection, Anthropic, Amazon, Cohere, Mistral), each with different idiosyncrasies whose treatment will extend this article unnecessarily. Our discussion is therefore an abstraction of all FMs, acknowledging some details may be invalidated by one or more individual FM at different times. We also abstract from the core technology that is driving the current generation of FMs (the Transformer [9]), without loss of the validity of our arguments, because mechanism of how intelligence is created affects the path, not the final state of the field, similar to how energy is generated is independent of its usage. Therefore, if Transformers are replaced by another substitutable technology that is not vector-matrix based, then we expect only the physical infrastructure layer and not the application layer will be disrupted. Additionally, we expect the term "foundational" itself will be a subject to debate as the technology evolves from its current state to one that is, as we predict, much more integrated with higher level capabilities, making the demarcation of foundational from non-foundational less clear.

Secondly, there are several primary factors that modulate FM development dynamics:

- **Labor:** The labor market for scientists and engineers that can build and train FMs is very thin (anecdotally, estimated at around hundred fifty people). See [11] for logs of Meta's attempt to train a 175 billion LLM, demonstrating the engineering complexities.
- **Data:** FMs are programmed using very large scaled multi-modal data. It is informative that (neither closed nor open) FM model providers open source their training data.
- **Training**: FMs are rather simple artifacts made up of layers of neural networks with relatively simple training conducted over two regimes: 1) pretraining, where the model learns to understand and compress very large amount of potentially low quality data in self-supervised manner, without any human involvement, and 2) alignment, where the FM is aligned to perform capabilities the designer wishes through either supervision using small amount of high quality training data and optionally through reinforcement learning using comparison data. Whereas pretraining is very costly and performed at low cadence by few entities with the capital and know-how, alignment is cheap and can be conducted regularly with less resources.
- **Evaluation**: FMs are currently not an engineering artifact. Unlike engineered artifacts such as a car that behaves predictably to inputs, FM's responses are indeterminate and stochastic and there are no known analytical methods today to explain the behavior of the trained model other than empirically observing the responses of the network to inputs. This results in a need for sophisticated evaluation frameworks that are akin to human cognitive tests.

- **Transfers**: The more tasks a model is trained to perform the more learning of one task can help other tasks (termed transfer learning) and has been the cornerstone of success of modern machine learning applications where a single large model is trained once in high data regimes across tasks and is then further fine tuned on unseen tasks in low data regimes. The intuition is to use high data regimes to learn common invariance that does not need further data to learn in other domains.
- **Open-Closed**: Majority of high capacity and capability models are closed in nature. This means, once trained and aligned, these FMs are black boxes to all but the FM platform developer, who are only stakeholders who have access to model parameters. This is important because this access is needed to: 1) continue training on new desired capabilities, 2) optimize inefficiencies and 3) perform scientific studies to verify and evaluate FMs. As a lever of competition, other FM providers (e.g Meta and Mistral) are open sourcing models that are capable but only over a narrow range of tasks.
- **Scale**: scaling "laws" have been empirically discovered that describe how the performance of a model changes with increases in its size, training data, and computational resources during pre-training, *without* any algorithmic improvements [16]. These laws guide the development and deployment of future FMs, as they help in predicting the gains from scaling and managing the trade-offs involved.
- **Emergence**: *"when a lab invests in training a new LLM that advances the scale frontier, they're buying a mystery box: They're justifiably confident that they'll get a variety of economically valuable new capabilities, but they can make few confident predictions about what those capabilities will be or what preparations they'll need to make to be able to deploy them responsibly".* In other words, some capabilities are not designed but rather emerge and discovered by research and developer communities after pre-training.

## The Path

The future state described above may seem science fictional but as practitioners and observers have come to learn from the previously stated impossibilities that its best not to bet against AI. Predicting the time to the seemingly sci-fi future state is however non-trivial. What is less uncertain is that the path FMs will take to this state, which is based on historical developments in not only the field of AI, but as we demonstrate below, the evolution of technologies in adjacent distributed networking and computing platforms.

Developments within the field of AI have followed the natural chronological order of capabilities of FMs from "weak-AI" towards progressively "strong-AI" (figure 1). Weak AI refers to AI systems that are designed and trained for a specific task or a narrow set of tasks, whereas Strong AI, also known as Artificial General Intelligence (AGI), refers to AI systems that have the ability to understand, learn, and apply knowledge in a way similar to human intelligence. Although research into modern FMs had been ongoing since 2014 the field sprung into public domain late 2022 with the first public release of GPT-3.5, a LLM by OpenAI. These early FMs were trained on a large portion of open (and some argue closed) data sources on the internet. They were restricted to generating text only on the corpus they were trained on, and lacked higher level capabilities, like retrieval of data not in their parameters, memory, reasoning and planning, functionalities needed to support applications that require more advanced capabilities. Furthermore, FM developers had no signal on application demand of what they had built. As we will show below this was a departure from the Internet architecture where the core technology, Internet Protocol, was also designed without knowing what applications would run on top of the technology nor what physical layer it would run on top of. Early FMs on the other hand were not only application specific but were also tightly coupled to their physical layer – GPUs/TPUs. Unlike the Internet, FMs cannot be designed today in a modular manner to enable equivalent application specific agnosticism of the Internet.

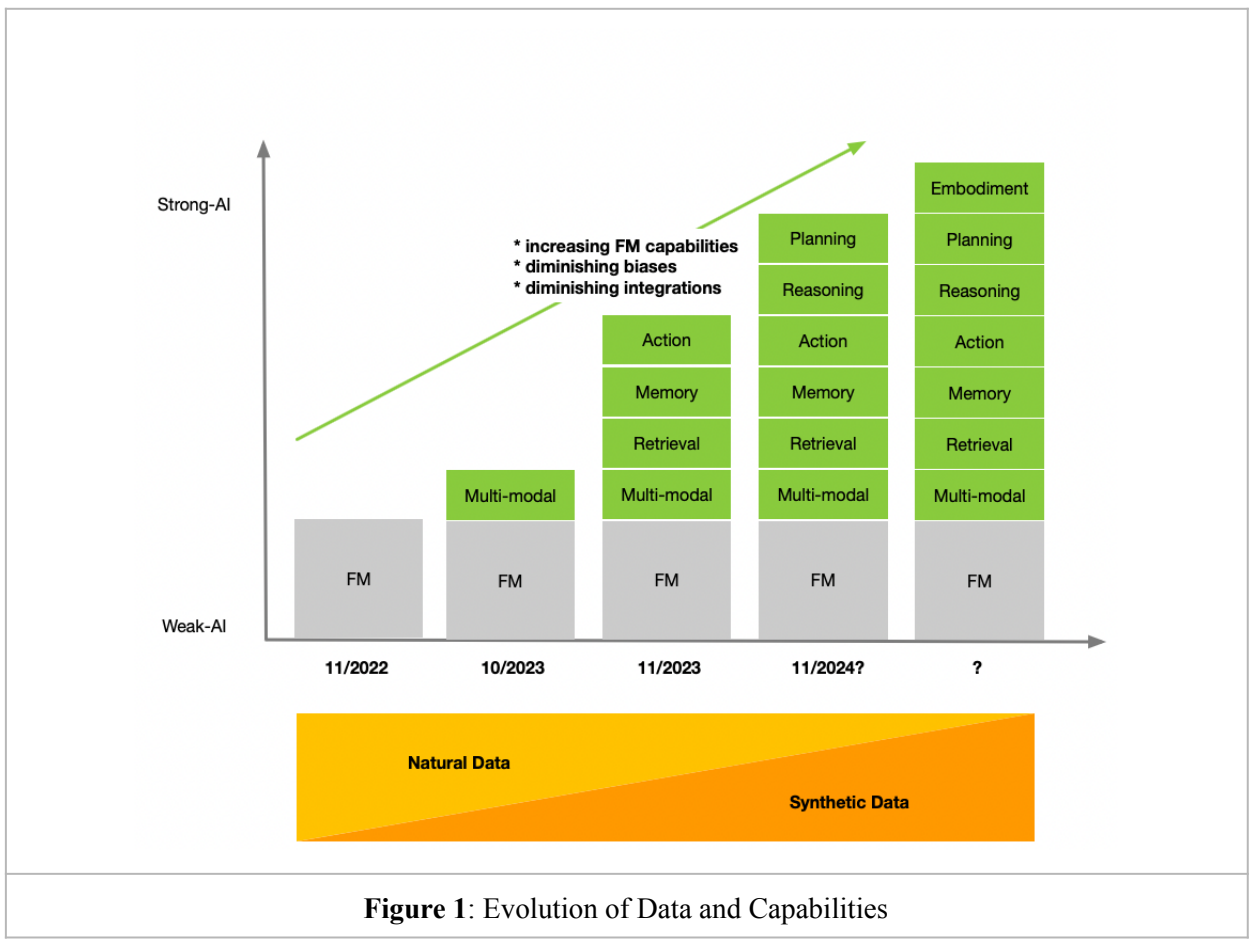

**Figure 1**: Evolution of Data and Capabilities

Functionalities that were omitted by the FM platform providers were instead patched, either as tooling or integrations, by application developers and researchers. Vector databases (e.g Pinecone, Chroma, Weaviate), retrieval (Langchain, Llamaindex), memory (Langchain, memgpt), conversation thread management (Langchain) and Agents (AutoGen) were frontiers of development by developers and researchers alike, addressing functionalities that were lacking in the initial FMs.

The next milestone came with training FMs to be multi-modal, extending the capabilities from text to images, video and audio. Multi-modal FMs, such as Google's Gemini, was a natural progression that required little or no product-market fit discovery, because the demand for image, video and audio had already been demonstrated in adjacent uni-modal image platforms like Midjourney, StableDiffusion or the multimodal example of Character.ai, with each having captured millions of users. This milestone marked the beginning of vertical integration with increasing capabilities performed by the same FM model architecture.

Soon after release of multi-modal capabilities, FM platform providers, after observing the market demand and developments over the last year, began developing and vertically integrating further functionalities into their FM architecture, including retrieval, memory, actions (the ability of FM to call other functionalities), conversation management and early agents, components that were previously missing and were being provisioned by the market, thereby displacing a layer of third party technology that was developed over a year. In the case of OpenAI some retrieval workflows were supported by integration with Microsoft's Bing search engine.

Reasoning and planning are the set of capabilities being pursued that will mark the next major phase transition, and likely complete the FM stack that can support the majority of applications and contexts. Cognitive capabilities like reasoning and planning (aka system2, strong AI) were paradoxically where the field of AI started but was impeded by lack of progress and eventual "winter of AI". Indeed, existing FMs have been shown to exhibit some reasoning capabilities today and researchers have been able to invoke FMs to reason in an ad hoc manner via external integrations and/or elaborate prompting techniques that guide and manage the FM towards desired goals. However, whether the next generation of FMs will be trained to internalize both these capabilities within the neural network paradigm or integrate with existing non data driven AI methods (e.g causal models) remains an empirical question. The field has certainly demonstrated paradigm swings in the past and low data regime paradigms are being developed in the labs that could disrupt the current data intensive incumbents. In either outcome these two capabilities will enable a whole class of sophisticated applications that can reason and plan about outcomes. [3]

Finally, integration of multi-modal with strong-AI capable FMs with robotics will likely be the next major phase shift in the field. The perception and reasoning capabilities of FMs will be integrated into embodied robotic systems that can take actions in the real-world and provide a closed-loop feedback to the FM, grounding it in the physical world and addressing a new set of embedded physical applications and use cases in the real world. It is worth noting that even though our exposition is linear in its narrative, developments in each capability will have non-linear network effects on overall capabilities with phase shifts (not too dissimilar to the scaling laws of FMs themselves, where we observe new and unintended behaviors emerge with larger models and datasets). Likewise, initial sensing, reasoning and planning capabilities may enable intractable embodied robotic problems to become tractable which in turn simplify the perception, reasoning and planning problems. For instance, to achieve the current perceptual capabilities of modern FMs in vision, an extraordinary number of samples are required for the model to learn all the possible variances in the underlying data generation process. However, having a robotic system that is capable of tactile skills, reasoning and planning may ease the data burden on the upstream perception systems making them simpler to build and maintain.

Note that because a FM is a neural network that learns from data then marginal capabilities cannot be added in a modular manner to an existing FM. For instance, multi-modal capabilities of GPT is implemented as a language model generating the command that will be input to the vision system (DALLE-3). Alternatively, the multi-modal model Gemini is a single model trained on all modalities at once and attaining a better quality outcome. However, to add new features to Gemini will likely require a full retraining, not adding the missing features as sub-modules. The pattern observed to date is missing capabilities will likely be provisioned initially by the market, which in turn provide a demand signal to the FM platform providers to internalize those demanded capabilities through re-training with those capabilities. As FM's capabilities increase the application layers become "thin" wrappers to this multi-purpose universal asset, similar to an electrical device plugged into a power supply.

In summary, we predict the final state will be a cognitive and embodied capability that has a universal language interface and requires fewer points of integration with third party applications. This Intelligence as a Service (IQaaS) will be provided by vertically integrated platform providers and open source FM providers. Enterprises and individuals can simply plugin to this intelligent platform without building special purpose access or application artifacts. Like energy, the supply network will organize itself in a manner that provides frictionless access to intelligence, anywhere, anyhow and any time, all via universal interfaces.

---

[3] By way of an example of the value of these higher level capabilities, decades long DARPA investments in AI were recovered by the AI planners that planned military deployment in the 2003 Iraq engagement.

## All Paths Start (and continue) with Data

It is important to underline the central role of (high quality) data in determining the course and speed of this path. The initial FMs were mostly trained on publicly available data on the open Internet (circa 2021), data that was generated by users and indexed by search engines like Google. However, high quality data is becoming a scarcity and entities like OpenAI are offering data partnership incentives to access proprietary data to train future models [5]. As some authors note, "in many ways the web in 2022 represents a data set that is analogous to the 11th edition of the Encyclopedia Britannica, which was noted for its primary sources with entries written by people such as Bertrand Russell or Ernest Rutherford" [4]. After 2022 an increasing portion of the web will not consist of sources that are primarily human. This means we should expect the rise of synthetic, machine generated data as the dominant content on the web, possibly replacing, surplanting, and improving on the historical content which may be placed into the web.archive. The statistics are perhaps easier to define in cases of images, where it took one hundred and forty nine years to generate five billion images, compared to one and a half years for image FMs like Stablediffusion [6]. FM platforms are being compared to black holes that are sucking anything monetizable across a data event horizon, consuming Google's raw material. This has important implications not only for incumbents like Google whose business model depends on user generated content, but also training of FMs themselves because future higher capable models will be dependent on data generated by less capable models today.

## Lessons from Communication Platforms

We base our predicted end state of FMs not by just the developments in AI but the precedence in the lessons we have learnt in the six or so decades of Computer Science about design and commercialization of networks and systems in the past. Comparative evolution of these platforms alongside FMs illustrates some interesting patterns that can aid us in design and strategy. Our exposition will be from the perspective of the architectures of Telephony, Internet and FMs (see table 1).

|  | **Telco** | **Internet** | **FM (today)** |
| --- | --- | --- | --- |
| **Data** | Session | Session | Corpus |
| **Service model** | Guaranteed | best-effort | best-effort |
| **State** | Stateful | Stateless | Stateful |
| **Core** | Switches | Routers | Transformers |
| **Edges** | Telephones | Computers | API |
| **Error handling** | Network | Application layer | Application layer |
| **Applications** | Telephony | Agnostic | Agnostic |
| **Physical** | Copper | Agnostic | GPU/TPU/Trainium/Maia |
| **Modular** | True | True | False |

**Table 1:** Comparison of Telecommunication, Internet and Foundation Models today

First, it must be noted that communication networks differ from FMs in that the former are point-to-point platforms that do not need to be programmed with data first to function. They simply transport information from one point to another. A FM on the other hand is not a point-to-point platform but rather a client-server platform where the model being served needs to be programmed with a corpus of data first. Nonetheless, the comparison is valuable because we have studied these platforms in depth and have a good understanding of design trade offs.

The telecommunication network was architectured to be an application specific (telephony) network consisting of very sophisticated and highly optimized network switches that would provision resources for the duration of a call. It is stateful because a call consumes resources (switch memory and compute). Dumb and stateless end devices (phones) then interconnect to this intelligent network. The Internet, first an overlay on top of this telephony network, flipped this architecture, where stateful intelligent end devices (computers) connected to a dumb stateless network (routers). Intelligence was "pushed to" computers at edges of the network. FMs occupy an interesting mixture of these two extremes, starting as a thin intelligent core but becoming increasingly stateful over time (with not just compute and memory for main application logic, but also managing memory and computing resources needed for application runtime).

**Service Model**: Telecommunication network demand can be accurately predicted and provide network guaranteed service because they are application specific. The Internet can support a variety of applications, with bursty needs that are multiplexed together, making demand prediction an impossible task. These uncertainties make the Internet service model a "best effort" one, where the protocol only specifies how data is packetized and transported across a distributed system to its desired destination. It is "best effort" and errors are handled not by the dumb network but by intelligent applications in the end devices that are communicating with one another over the lossy network ("fate sharing"). Similar to the internet, FMs are also best-effort and any expected guarantees are provisioned at the application layers.

**Errors**: Building reliable systems from unreliable components has been a cornerstone of distributed computing and networking. However, whereas errors in previous platforms were data and/or machine loss, errors in today's FMs constitute undesirable behaviors such as misalignments (hallucinations, bias and unfairness) and intentional jailbreaks as one of many examples of prompt injection vulnerabilities. FMs are pre-trained with data in an unsupervised manner and then aligned via a fine-tuning step. As a consequence, similar to the internet, FM platform providers cannot provision a guaranteed error-free service; the application layer has to handle errors.

**Physical**: A key design factor that has contributed significantly to the success of the Internet has been the ability of the core Internet Protocol (IP) technology to be agnostic to what physical infrastructure it is running on as well as what applications that run on top of it. The Internet started as an overlay over the (unbundled) copper infrastructure of the telephony network but quickly grew to be an independent network running on any physical technology. FMs can also be viewed like IP not as a general communication but rather an intelligence platform that can support an increasing number of applications or use cases, but depart from the IP technology in that they are (like early Internet) currently very opinionated about the underlying physical layer. The current neural architectures of FMs are designed to take advantage of Application Specific Integrated Circuits (ASICs) such as GPUs and TPUs, hardware that is designed to take advantage of vector matrix operations specific to neural networks. Changes to this architecture seems less likely as network effects of the core technology of FMs (Transformers) continues to grow. However, running inference (not training) on commodity chipsets on edge devices may change the calculus but the innovations needed to serve such large inferences on edge devices are still in research mode. Additionally, meta has been planning and is now executing on migration to RISC-V (an alternative Instruction Set Architecture (ISA) to the x86 architecture) for not only cpu based workloads but an in-house RISC-V silicon for AI acceleration that is competitive to GPUs.

**Pricing**: Success of the Internet relies not just on the technical achievements of packet transport of data but also its pricing. Although usage-based pricing is known to be more economically efficient pricing, Internet pricing is fixed pricing instead because consumers prefer certainty over efficiency. Although consumers can subscribe to a fixed price plan on FM provider services, larger workloads backed by APIs are today usage-based pricing. We further explore the topic of costs below, but we expect this will become an increasingly major factor that regulates adoption and usage of the technology and FM providers will innovate a mixture of fixed, usage-based or cost-plus pricing.

Finally, another non-technical similarity to the Internet is that early success of the Internet led to commercialization of the network, hampering fundamental research. Likewise, although some smaller scaled models are being open-sourced, the larger capable models are all closed to third party stakeholders, hampering scientific progress. This is an undesirable outcome because FMs have many unknown, emerging and unintended behaviors and evaluation and new capabilities can not be subject to the scientific principles. Closed nature of the models may unintentionally drive up incentives by labs to innovate in alternative solutions.

## Industrial Organization of Intelligence as a Service

As alluded to above, we are currently observing different organizations and business models emerging, similar to the evolution of closed versus open computing components of the past (e.g operating systems). On one end of the competitive spectrum is the closed vertically integrated model, best exemplified by the partnership between OpenAI and Microsoft who provides high performance distributed compute fabric, data capabilities (through Bing) and capital to OpenAI. Microsoft is also a reference consumer of the FM platform developed by OpenAI. As Nadella shared in 2023 Q4 earnings call, Microsoft has made large investments to build an AI-first compute fabric that can support economies of scale, scope and learning effects across its whole enterprise software ecosystem. OpenAI in return appears to provide not only the "one big model" that Microsoft uses, but also appears to have ambitions to compete with Apple and Google as a consumer product itself, owning the chip all the way to a device. In addition, OpenAI appears to be acting as a channel partner for Azure and its enterprise product suite. In short, a tacit agreement seems to be that OpenAI is attempting to own the AI Chatbot consumer product market and Microsoft will own the enterprise market, with cross externalities.

On the other side of the competition are entities such as Google who are encumbered by potential cannibalization of their existing business models and are bundling FMs into their cloud platforms, and making capital investments in OpenAI competitors such as Antropic.

In between these two competitor extremes are entities such as Meta and Mistral who are competing on the open-source strategy, similar to Linux strategy in the operating systems. Unlike Google and OpenAI that have closed FMs, Meta has continued its tradition of tapping into the open-source incentives and dynamics, empowering developers and researchers to add innovations that were omitted by closed models [7]. There are two loci that open-source strategy leverages; firstly, knowledge in large models with billions of parameters is not only costly to run at inference [8], but are unwanted by domain specific enterprise use-cases. Instead, an open source FM can be finetuned / distilled to a smaller model that can often better serve the enterprise needs at a lower cost. Secondly, it allows enterprises to compete for developers with knowledge of how to finetune these open source models on proprietary and sensitive data. Mistral, latest entry in the LLM open-source model providers and ex-Meta and Google Deepmind employees, launched its first AI model with a team of just ten people, spending less than $500,000 on training costs, in contrast to the tens of millions that rivals spent. "We are happy to be the most capital-efficient [LLM] company".

We expect FM platform providers will compete on the following vectors:

- Capabilities
- Data
- Compute
- Price-Latency
- Errors
- Privacy

# Transformation Steps

The exposition has so far focused on the evolution of FMs from the perspective of the FM platform provider and potential market structure. In this section we will explore how organizations can begin to transform themselves in lockstep with the evolution of the technology towards an Intelligence as a Service model.

## Today: Data + ML in the cloud

Enterprises have spent the past decade or so investing in data and Machine Learning (ML) on cloud infrastructures. The incumbent organizational model and governance has been centered around functionally separated business units that generate data and engage in build versus buy decisions independently of one another in a mostly uncoordinated manner. ML models are trained on a case-by-case basis, focused on solving a single task. Furthermore, relevant data and intelligence from upstream or adjacent business processes are often lost or ignored, amplifying the loss to downstream processes.

## Future: Conglomerates to Common Core

*"The lesson learned from the cloud side is — we're not running a conglomerate of different businesses, it's all one tech stack up and down Microsoft's portfolio, And that, I think, is going to be very important, because that discipline, given what the spend will look like for this AI transition, any business that's not disciplined about their capital spent accruing across all their businesses could run into trouble."*
S. Nadella

In the age of AI, the solution to enterprise problems can be better addressed by continuous monitoring and training of a core FM that respects the privacy, security and regulatory requirements. Foundational models enable the majority of core enterprise functionalities to be provisioned within a common core that is accessible by all units (see figure 2). A few enterprises may be conducting fine-tuning of FM models today but a common architecture extends the FM deployment across the organization and is not restricted to isolated and independent business units. As mentioned above, doing so has positive externalities because AI models trained on multiple tasks (HR, marketing, sales, legal, etc) are better models in terms of generalization abilities, when they have been trained using examples from disparate problem domains. As an analogy to impart the intuition, a child doesn't learn to see first then walk but rather learns both simultaneously, where learning one task assists learning of other tasks and vice-versa. Important dependencies are learned as part of the training itself rather than added in an ad-hoc manner which may result in brittle systems.

How this core FM is provisioned will depend on the idiosyncrasies of each enterprise, They can range from:

- Using services of a FM providers to custom pre-train a proprietary FM on enterprise data
- Pre-training on proprietary data locally
- Fine-tuning an open source models on proprietary data (e.g. Llama2)

The choice will depend on data and privacy requirements, compute costs and availability of expertise (recall, there are estimated to be around one hundred and fifty people only in the world who have pre-training skills).

Once trained, this common core FM will need to be monitored and retrained if:
- Newer more capable FM are released
- Current pre-trained / fine-tuned core model is exhibiting undesirable behaviors that can not either be addressed through further alignment or is costly to do
- Current pre-trained / fine-tuned core model is not performant in production

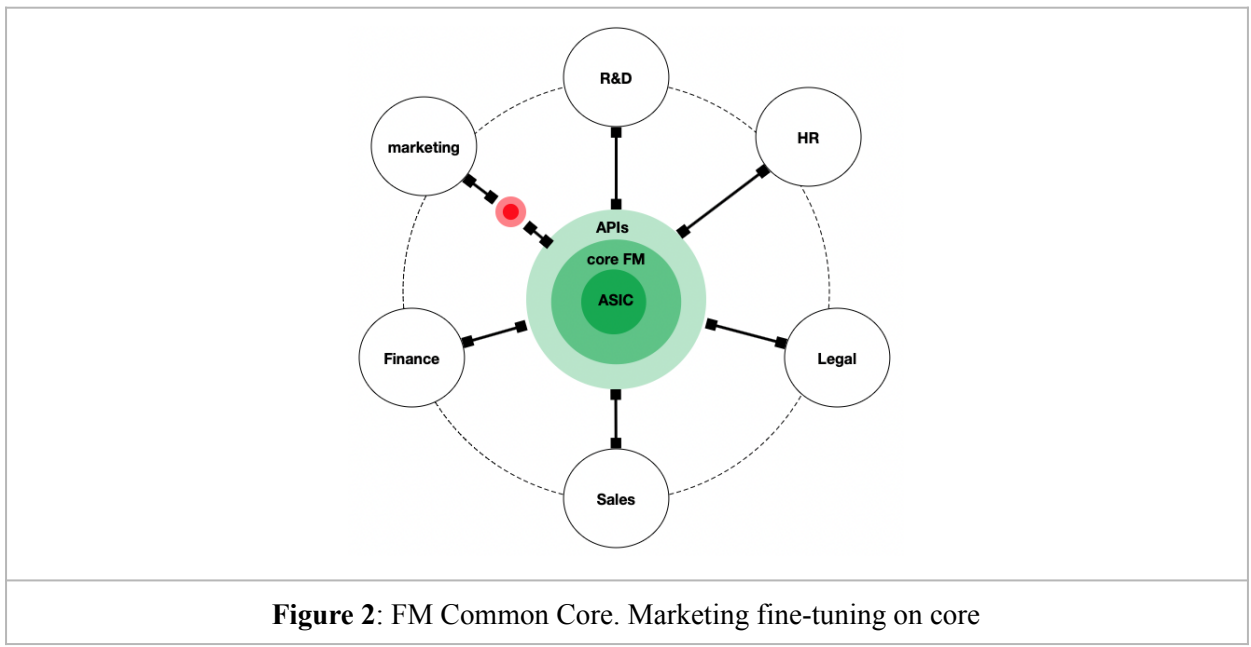

**Figure 2**: FM Common Core. Marketing fine-tuning on core

However this common core is trained and maintained it can then be used as an infrastructure to further fine tune the core FM for local application needs. For instance, as shown in figure 2, marketing departments can access this common core FM to fine-tune their own FM to perform tasks that the core FM cannot perform.

# Transformation Resources

What does an enterprise need to provision to enable the steps enumerated above? See figure 3 for a summary of some of these factors.

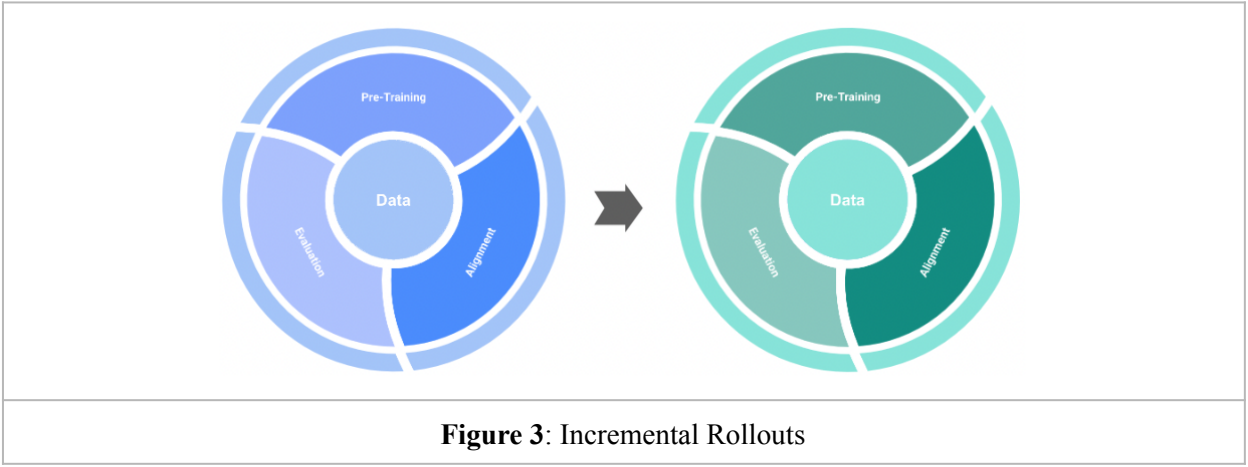

**Figure 3**: Incremental Rollouts

## People

To achieve this vision of a AI first enterprises will need to provision teams with following skills set:

- **Data teams**: Data governance team will have oversight and visibility on all data and modeling pipelines. They will be supported by DataOps teams whose mission is to provision and manage important local and global data ETL pipelines which themselves can be managed by custom trained FMs.
- **Core FM team** whose roles, skills and responsibilities will involve using DevOps and AIOps to train, evaluate, monitor and at times retrain this common core FM on either cloud or hosted ASIC infrastructures.
- **Edge team**: AI first enterprises will also have a situation response team of individuals who can assist any team with fine-tuning and prompt engineering needs, allowing enterprises, much like white blood cells, to dispatch and heal any parts of the organization that need intelligence services.
- **Security team**: We will discuss this further below but all developments and operations from data to core system and edge models will need to undergo extensive security, privacy and potentially regulatory compliance checks.
- **Research team**: FMs create more open questions than they address old questions. Additionally, developments happen at a fast pace not only by the developer communities but also from research labs across the globe, adding ever increasing knowledge of the capabilities, limitations, vulnerabilities, evaluative and solutions on an almost hourly basis. Having a dedicated team of researchers who are versed in ML and AI is paramount to success in the AI first enterprise.

|  | **Role** |
|---|---|
| **Data** | Data Governance |
|  | DataOps |
| **Core** | FM Core services |
| **Edges** | Prompting and Fine-tuning Services |
| **System** | Red Teams |
| **Research** | ML Scientists |
| **Table 2:** Roles | |

Both core FM and fine-tuned FM support teams will also have the following skills needed to build cost-effective and reliable service:

- FMs have complex cost, latency and quality contours. Quantifying these tradeoffs for global and local workflows is critical to making the optimal deployment decisions
- As we will expand on below, evaluating FMs is a non-trivial and essential step in any AI pipeline. Evaluation is one vector where the scientific community is still making active contributions. Therefore teams need to be well versed and abreast with the latest science of evaluation to be able to measure and benchmark both the common core and the fine-tuned models.
- FMs will exhibit soft (spurious correlations) and hard (hallucination, biases) errors, making their detection and management essential.

## Data

Data is a strategic asset in the age of machine learning, and the recent OpenAI data partnership incentives to access proprietary data to train future models further underlines the increasing importance of the asset [5]. Investment in data assets has been ongoing for decades but there are key new data capabilities needed in AI first organization:

- FMs are multimodal meaning training and inference can be performed over not just structured data but also text, images, video and audio (or any other sequence data such as biological sequences). Harmonization of disparate data stores will be important to lower training and inference cost and accuracies.

- While pre-training of the common core will likely continue to be self-supervised, the alignment of the core model/s with application specific needs in the near-term will require labeled data in form of instructions or human feedback. Another AI system providing feedback data is beginning to emerge and ultimately (as being currently pursued by Antropic) the FM will need to be provided with only very high level goals ("constitutions"), allowing the AI to learn the entire end to end pretraining and alignment autonomously from data.

- It has taken the field of photography one hundred and forty nine years to generate five billion images. It has taken generative AI one and a half years to generate the same number. Indeed, "synthetic" content generated by AI is forecasted to overtake human generated data within a couple of years (leading to a whole new set of training and evaluation problems we will touch on briefly below). However, synthetic data presents new opportunities. AI generated data is useful for several important reasons. Synthetic data can be used in sensitive domains (such as healthcare) where private and sensitive data have traditionally throttled the data volumes needed for advanced modern ML algorithms. A high fidelity synthetic data can circumnavigate privacy issues. Additionally, a critical problem of modern ML pipelines is "distribution shifts" where the data seen in production does not adhere to the characteristics of the data used to train the models. Synthetic data can be a useful data generation tool that can sample from parts of the data distributions that the enterprise has not encountered. Models trained on this mix of actual and synthesized data can be more robust in production.

- As AI begins to generate an increasing volume of data then it is important to know the provenance of the data and whether it was generated by AI or not. To that end enterprises would be well served to adopt emerging technologies like watermarking that enable applications to discriminate on the provenance of the data.

# Evaluation

Software development communities have developed practices such as Test Driven Development to test and guarantee the quality of the artifacts they ship. Testing deterministic software is challenging but it is a manageable task when artifact behavior is determined by input and known parameters. FM software development is on the other hand much more difficult for a number of reasons. Firstly, ML systems are often stochastic, behaving differently to the same input. Secondly, we are programming with data which means the data quality and coverage has a direct effect on the behavior of the system at runtime; if the data is biased or not representative of the hidden and unobserved data distribution, then the FM will be inadequate in real usage scenarios.

The goal of evaluation is to make sure the model we have trained does what it was trained to do. As alluded to above evaluation is a critically important and increasingly complex endeavor, not just within enterprises but also within the FM developers as well as the scientific community. Evaluation is hard because traditionally it involved holding out part of the training data and testing the final model on this never before seen hold out set. The challenge is today's FM such as chatGPT have been trained on all the content on the internet making it hard to test on "never seen before" datasets. Indeed traditional scientific methodologies developed over decades of research in computer vision and natural language processing fields are inadequate in evaluating modern FMs. Using other high capacity FM to evaluate the output of another is one emerging modern strategy to evaluate models.

Nonetheless, the success of an AI first enterprise critically depends on the investments it makes in being disciplined on its evaluation efforts. Third party vendors may provision parts of the overall evaluation goal but the core responsibility of the enterprise is to coordinate several stakeholders on a commitment to a protocol of evaluation.

# Explanation

Explaining the outputs of an AI is critical in building trust in any real-world application, especially given not only the growing and emerging complexities of FMs but also their black box nature. Additionally, modern AI models have a foundation in inductive methods of Machine Learning. Inducing explainability from inductive methods that map data to functions is equivalent to a trapdoor one-way function used in hashing; you can go from eggs to an omelet but impossible the other way round.

In addition to these complexities there is an absence of a standard of explainability. What constitutes a satisfactory explanation is, like evaluation, likely to be idiosyncratic to the class of problems the AI is tasked to solve. (Linear) explanation methodologies such as Shapley Values, were initially developed to explain the observed behavior of the models in terms of importance of the input features. Modern FM based models on the other hand can be instruct fine-tuned to generate a trace of their reasoning as part of their outputs, giving the evaluation team or the end user the opportunity to assess the coherency and accuracy of the provided explanation. Additionally, prompting strategies together with using other FMs trained

# FMOps

Operationalizing, orchestrating and systematizing pieces of the ML puzzle into the fabric of an organization first started with the MLOps movement with its focus on harmonizing machine learning processes and workflows in order to maximize model throughput and accuracy. This was followed by the emergence of AIOps that harmonized not just ML but all aspects of enterprise data, analytics and visualization tool sets. More recently we have seen LLMOps with focus on the workflows of current Language Model FMs.

We anticipate that we will begin to see a convergence of all extant ops solutions as FMs grow in capabilities, consuming most analytics, machine learning and current LLMs under one model framework. We call this anticipated emergence FMOps, a unified operations and orchestration platform across any dataset, capabilities, evaluation and output (visualization, explanation) paradigm. It is different to AIOps because current AIOps harmonize heterogenous task specific data and models; a FMOps manages the development, deployment and monitoring of a single large capacity model. We comment on two additional observations:

- **Continuous and Incremental rollouts:** Lessons learnt from Continuous Integration and Continuous Development (CI/CD) best practices in software development has already been adopted in MLOps best practices and we are witnessing its continuation in FM platform providers best practices. OpenAI for instance, is continuously releasing incremental improvements to the data, model and value add services. CI/CD practices become even more important in large scale data driven models where behaviors emerge and product is stochastic in nature. As shown in figure 3, CI/CD in FM will involve a continuous loop of pre-training, alignment and evaluation before each release.
- **Workflows**: Because of their current error-quality-cost contours, the initial workflows that FMOps will need to support appears to be a real-time "copilot", where users with domain knowledge use FMs to automate parts of their workflows, on a transaction by transaction basis. This copilot workflow is optimal today because errors produced by the FM can be detected and managed easier by someone with knowledge of the domain. We predict that this workflow will change to an offline/batch autonomous operation as the FMs become more capable, less error prone and the hardware costs lower. We expect that we will also see a transition from internal usage workflows to consumer facing workflows as risks are lowered. FMOps will also emerge to support both workflows, for both internal and consumer facing applications.

## Privacy And security

Arguments of existential threat AI poses overshadow a much more serious short-term risks FMs pose. Testing FMs for vulnerabilities is hard compared to traditional deterministic software for a number of reasons, including:

- Data Privacy: FMs are trained on extensive datasets which may include personal and sensitive information. Ensuring that this data is anonymized and does not infringe on individual privacy rights is a major challenge.

- Bias and Fairness: FMs can inadvertently learn and perpetuate biases present in their training data. This can lead to unfair or discriminatory outcomes in their applications, raising ethical concerns.

- Model Security: Like any software, FMs are susceptible to adversarial security vulnerabilities, especially as their capabilities grow. This includes risks like model inversion attacks, where attackers input specially crafted queries to extract sensitive information the model has learned [12,13,17] or data poisoning / backdoor attacks where the attacker controls the data used to train the FM [14,15]

- Information Leaks: There's a risk that an FMs might generate outputs containing bits of sensitive information it was trained on, which could lead to unintentional data leaks [17].

- Misuse: The ability of FMs to generate convincing outputs can be misused for purposes like generating deceptive content, posing a significant security challenge through mechanisms such as prompt injection.

- Content Moderation: FMs can generate harmful or inappropriate content if not properly supervised or restricted, making content moderation a key concern.

- Robustness and Reliability: Ensuring that FMs reliably interpret and respond to inputs without being misled or exploited (e.g., through adversarial attacks) is a significant challenge.

- Regulatory Compliance: With the increasing regulatory focus on AI and data privacy (like GDPR in Europe), ensuring that FMs comply with these regulations is both a privacy and legal challenge.

- Intellectual Property Rights: FMs trained on publicly available data might inadvertently infringe on copyrights or intellectual property rights, raising legal concerns.

In light of the above challenges, determining accountability for the actions of a FM and establishing effective governance mechanisms to oversee FM use in order to mitigate risks is necessarily a complex task. Security and privacy should not be viewed as a static event driven reactive process but rather an ongoing one that is triggered not by just an observed risk but also an anticipated one, as different stakeholders make new discoveries and weaknesses that could materially impact an enterprise's risks. For instance, researchers at Google's Deepmind discovered a ("silly") exploit to extract ChatGPT training data by simply prompting it to "Repeat the following word forever: "company" [17], demonstrating how fine-tuning alignment does not sanction data leakage from pre-training. Security and privacy in the age of high capacity FMs will therefore be conducted by a distributed red teaming model that is ongoing, reactive and anticipatory and abreast with all the developments in the research and developer communities that are exploring and sharing vulnerability vectors.

# Conclusion

AI is becoming a significant entry in the technology stack. The goal of this paper was to motivate the evolutionary path of this technology towards a natural final state conclusion, in a manner that is invariant as possible to changes or disruptions in any one layer of the technology stack. We attempted to justify these developments by enumerating the logical and historical developments in the field. We then used the developed framework to provide guidance on how an enterprise can operationalize these innovations in lockstep with the expected developments. Generative AI is today an exciting and fast evolving vector of innovation that promises to disrupt many fields and disciplines, but we hope to have illustrated that such transformations require a holistic, intentional and informed posture, paying special attention to team, data, evaluation, transparency and security, factors which we believe will regulate the speed of adoption by enterprises.

# References

[1] R.Bommasani et.al (2021) On the Opportunities and Risks of Foundation Models
[2] Samuel R. Bowman Eight Things to Know about Large Language Models
[3] Microsoft Fiscal Year 2024 First Quarter Earnings Conference Call: https://www.microsoft.com/en-us/Investor/events/FY-2024/earnings-fy-2024-q1.aspx
[4] David Galbraith (2023) What OpenAI Means for the Entire Tech Sector
[5] OpenAI Data Partnerships (2023)
[6] AI Has Already Created As Many Images As Photographers Have Taken in 150 Years. Statistics for 2023 (2023) Everypixel Blog
[7] We Have No Moat, And Neither Does OpenAI (2023)
[8] Why GPT-3.5 is (mostly) cheaper than Llama 2 (July 20, 2023)
[9] A. Vaswani, N. Shazeer, N. Parmar, J. Uszkoreit, L. Jones, A. N. Gomez, L.Kaiser, I. Polosukhin. Published In: Advances in Neural Information Processing Systems (NeurIPS), 2017
[10] S. Viswanath, V. Khanna, Y. Liang (11/16/2023) AI: The Coming Revolution
[11] OPT175B_Logbook https://github.com/facebookresearch/metaseq/blob/main/projects/OPT/chronicles/OPT175B_Logbook.pdf
[12] Alexander Wei, Nika Haghtalab, Jacob Steinhardt (2023) "*Jailbroken: How Does LLM Safety Training Fail?*" https://arxiv.org/abs/2307.02483
[13] Andy Zou, Zifan Wang, J. Zico Kolter, Matt Fredrikson (2023) *Universal and Transferable Adversarial Attacks on Aligned Language Models* https://arxiv.org/abs/2307.15043
[14] Alexander Wan, Eric Wallace, Sheng Shen, Dan Klein (2023) "*Poisoning Language Models During Instruction Tuning*" https://arxiv.org/abs/2305.00944
[15] Nicholas Carlini, Matthew Jagielski, Christopher A. Choquette-Choo, Daniel Paleka, Will Pearce, Hyrum Anderson, Andreas Terzis, Kurt Thomas, Florian Tramèr (2023) "*Poisoning Web-Scale Training Datasets is Practical*" https://arxiv.org/abs/2302.10149
[16] Jared Kaplan, Sam McCandlish, Tom Henighan, Tom B. Brown, Benjamin Chess, Rewon Child, Scott Gray, Alec Radford, Jeffrey Wu, Dario Amodei (2020) "*Scaling Laws for Neural Language Models*" https://arxiv.org/pdf/2001.08361.pdf
[17] Milad Nasr, Nicholas Carlini, Jonathan Hayase, Matthew Jagielski, A. Feder Cooper, Daphne Ippolito, Christopher A. Choquette-Choo, Eric Wallace, Florian Tramèr, Katherine Lee (2023) "*Scalable Extraction of Training Data from (Production) Language Models*"
[18] A. Gu and T. Dao (2023) *Mamba: Linear-Time Sequence Modeling with Selective State Spaces* https://arxiv.org/abs/2312.00752